\documentclass[12pt,english]{paper}

\usepackage[T1]{fontenc}
\usepackage[latin9]{inputenc}
\usepackage{textcomp}
\usepackage{amstext}
\usepackage{graphicx}
\usepackage[authoryear]{natbib}
\usepackage{babel}



\begin{document}

\title{Voltage clamp analysis of nonlinear dendritic properties in prepositus hypoglossi neurons}

\author{\small{Christophe Magnani, Daniel Eug\`{e}ne, Erwin Idoux and L.E. Moore{*}}}

\institution{CESEM - UMR8194 - CNRS - Universit\'{e} Paris Descartes}

\maketitle

\small{Address : 45 rue des Saints-P\`{e}res, 75270 PARIS, FRANCE - Tel.:33-(0)142863398,
Fax:33-(0)142863399}


{*}Corresponding author

\begin{abstract}
The nonlinear properties of the dendrites in prepositus hypoglossi
neurons are involved in maintenance of eye position. The biophysical
properties of these neurons are essential for the operation of the
vestibular neural integrator that converts a head velocity signal
to one that controls eye position. A novel method named QSA (quadratic
sinusoidal analysis) for voltage clamped neurons was used to quantify
nonlinear responses that are dominated by dendrites. The voltage clamp
currents were measured at harmonic and interactive frequencies using
specific stimulation frequencies, which act as frequency probes of
the intrinsic nonlinear neuronal behavior. These responses to paired
frequencies form a matrix that can be reduced by eigendecomposition
to provide a very compact piecewise quadratic analysis at different
membrane potentials that otherwise is usually described by complex
differential equations involving a large numbers of parameters and
dendritic compartments. Moreover, the QSA matrix can be interpolated
to capture most of the nonlinear neuronal behavior like a Volterra
kernel. The interpolated quadratic functions of the two major prepositus
hypoglossi neurons, namely type B and D, are strikingly different.
A major part of the nonlinear responses is due to the persistent sodium
conductance, which appears to be essential for sustained nonlinear
effects induced by NMDA activation and thus would be critical for
the operation of the neural integrator. Finally, the dominance of
the nonlinear responses by the dendrites supports the hypothesis that
persistent sodium conductance channels and NMDA receptors act synergistically
to dynamically control the influence of individual synaptic inputs
on network behavior.
\end{abstract}

\section*{Introduction}

Mathematical models based on the experimentally measured biophysical
properties of neurons generally consist of complicated sets of differential
equations derived from the historical \citeauthor{Hodgkin1952a} (1952)
model. Extending the HH formalism to branching neurons requires a
large number of parameters that must be determined to obtain a realistic
neuronal model. The techniques previously employed to measure these
parameters involve either linear admittance (or impedance) measurements
or \emph{ad hoc} extrapolations from voltage clamp experiments with
poor space clamp control. Thus, it is important to consider more refined
theories from nonlinear analysis, such as nonlinear dynamics of neurons
(\citealp{Gutkin1998}; \citealp{Izhikevich2002}) or nonlinear system
identification (\citealp{Marmarelis1973c}; \citealp{French1976};
\citealp{Victor1980}; \citealp{Boyd1983}; \citealp{Schetzen2006}).
A goal of nonlinear analysis is not just a refinement of the linear
systems approach, but the development of a fundamental insight into
how neurons process information.

In a seminal paper, Fitzhugh derived the equations of the nonlinear
response for a single sinusoidal voltage clamp (\citealp{FitzHugh1983}).
This approach has been extended to the quadratic response for a multi-sinusoidal
voltage clamp and been developed as a matrix theory termed quadratic
sinusoidal analysis (QSA) (\citealp{Magnani2010}). The proposed experimental
approach is essentially based on QSA, requiring that the stimulus
amplitudes evoke mainly linear and quadratic responses. In addition,
QSA provides the mathematical tools for a model independent analysis
of quadratic nonlinearities and provides an innovative way to quantitatively
describe real neurons and their models. The measurement of nonlinearities
in neurons under normal physiological conditions is clearly important
in order to understand how they process synaptic inputs, which typically
evoke 5-10 mV post-synaptic nonlinear responses.

Two types of neurons of the rat prepositus hypoglossi nucleus (PHN)
were investigated. Both types clearly manifest nonlinearities at multiple
subthreshold step levels. The type D neurons are known to show marked
spontaneous, voltage dependent and irregular oscillatory properties.
By contrast, type B neurons, the majority in this nucleus, are non-oscillatory
and have regular spontaneous activity that is highly dependent on
a significant persistent sodium (gNaP) conductance (\citealp{Vervaeke2006}).
In this paper, the novel QSA method has been used to investigate the
quadratic response to time varying voltage clamped stimuli and establish
a quantitative characterization of the nonlinear behavior in order
to understand neuronal responses elicited by normal physiological
synaptic inputs.

It will be shown that at physiological levels of stimulation, neurons
and their models can generate significant responses at harmonic and
interactive frequencies that are not present in the input signal.
Thus, the nonlinear frequency responses contain more frequencies over
a wider frequency band than the input signal. As a consequence they
provide significant amplification at dynamically changing membrane
potentials. The use of stimuli with multiple input frequencies allows
one to probe neuronal function and characterize it by a matrix of
quadratic interactions, namely the QSA matrix. It is then possible
to extract information about active membrane properties from this
matrix by eigendecomposition. Finally, biologically realistic simulations
have been implemented using neuronal models based on vestibular neuronal
experimental data. These simulations suggest that the nonlinear responses
in voltage clamp are dominated by active dendritic structures.

\section*{Materials and Methods}

\subsection*{Whole-cell patch-clamp recordings and statistical analysis}

This paper is both a theoretical and experimental nonlinear approach
to neuronal function that adds to previous steady state linear analyses
(\citealp{Fishman1977b}; \citealp{Murphey1995}). It provides a quantitative
assessment of quadratic responses of both data recorded from individual
neurons and their corresponding biophysical models. Experiments were
carried out on male Wistar rats (25- to 52-days-old) supplied by Centre
d'Elevage Roger Janvier (Le Genest Saint Isle, France). All efforts
were made to minimize animal suffering as well as the number of animals
used. All experiments followed the guidelines on the ethical use of
animals from the European Communities Council Directive of 24 November
1986 (86/609/EEC). Brain dissections were performed as described elsewhere
(Idoux et al., 2008). Briefly, after decapitation under deep anesthesia,
the brain was quickly removed and placed in ice-cold, phosphate/bicarbonate-buffered
artificial cerebro-spinal fluid (ACSF), which included (in mM) 225
sucrose, 5 KCl, 1 NaH$_{\text{}2}$PO$_{\text{4}}$, 26 NaHCO$_{\text{3}}$,
0.25 CaCl$_{\text{2}}$, 1.3 MgCl$_{\text{2}}$, 11 glucose and was
bubbled with 95\% O2-5\% CO$_{\text{2}}$ (pH 7.4). Four or five 250
\textmu{}m thick, coronal slices containing the PHN were cut from
the brainstem with a microslicer (Leica, Rueil-Malmaison, France)
and transferred into an incubating vial filled with a regular ACSF
containing (in mM) 124 NaCl, 5 KCl, 1 NaH$_{\text{2}}$PO$_{4}$,
26 NaHCO$_{\text{3}}$, 2.5 CaCl$_{2}$, 1.3 MgCl$_{\text{2}}$, 11
glucose and bubbled with 95\% O$_{\text{2}}$ and 5\% CO$_{\text{2}}$
(pH 7.4). Slices were then placed one at a time in the recording chamber
maintained at 32-34\textdegree{}C, where the slice was superfused
with regular ACSF at a constant flow rate of 3 mL min-1.

Patch-clamp pipettes were pulled from borosilicate glass tubing to
a resistance of 5-8 M$\Omega$. The control internal solution
contained (in mM) 140 K-gluconate, 2 MgCl$_{\text{2}}$, 10 HEPES,
0.1 EGTA, 4 Na$_{\text{2}}$ATP, and 0.4 Na$_{\text{2}}$GTP (adjusted
to pH 7.3 with KOH). The junction potential for this internal solution was not subtracted for the potential measurements or the model simulations. PHN neurons were visualized with a Nomarski optic
microscope under infrared illumination. Recordings were made with
an Axoclamp 2B amplifier (Axon Instruments, Union City, CA, USA) or
a Multiclamp 700B (Molecular Devices, Sunnyvale, CA, USA). The spontaneous
discharge was first recorded in the current-clamp mode for 8 to 10
minutes once a stable level had been reached and the recorded PHN
neuron was determined as B or D type (see Idoux et al., 2008). PHN
neurons that had resting membrane potential more negative than -50
mV and a spike amplitude > 45 mV were selected for the voltage clamp
experiments. Types D and B neurons from the prepositus hypoglossi
nucleus were measured for different stimulus amplitudes and membrane
potentials. Based on the criteria for time invariance discussed in
the Rationale section, five type D and six type B neurons were selected
for detailed analysis. All measurements were made with stimuli applied
for twice the duration used in the analysis. Only the last half of
the record was used to assure that a steady state condition was reached.
At voltage clamp potentials near threshold, transient currents due
to uncontrolled action potentials occasionally occurred in the non-analyzed
initial part of the recording, however they were completely inactivated
by the maintained depolarization with no firing during the latter
analyzed part of the record. In some experiments 25-50 $\mu$M
NMDA (Sigma, St Quentin Fallavier, France) was applied in the presence
or absence of 2 \textmu{}M TTX (Tocris, Bristol, UK).

The data acquisition was done with a PC-compatible computer running
Windows XP, using MATLAB scripts (MATLAB 7.0, MathWorks, Natick, MA,
USA). Recordings were low-pass filtered at 2 kHz and digitized at
5 kHz (BNC-2090 + PCI-6052E, National Instruments, Austin, TX, USA).
The OneSidedPValue (p-value) of MeanTest{[}x-y{]} for paired differences,
mean values and standard deviations (\textpm{}SD) were calculated
with the HypothesisTesting package of MATHEMATICA 7.0, (Wolfram Research,
Champain, IL, USA). 

\subsection*{Rationale}

PHN neurons were analyzed with QSA, which specifically selects harmonic
and intermodulation frequencies, described as follows. If a double
sinusoidal input has frequencies $f_{1}$ and $f_{2}$ then the linear
response will have exactly the same frequencies $f_{1}$ and $f_{2}$.
However, the quadratic response will include additional harmonics
$2f_{1}$ and $2f_{2}$ as well as intermodulation products $f_{1}+f_{2}$
and $\left|f_{1}-f_{2}\right|$. This principle can be generalized
to a multi-sinusoidal input in which case the quadratic response will
include double of each input frequency as well as sum and difference
of each pair of distinct input frequencies. A quadratic response can
generate frequency overlaps when distinct combinations of input frequencies
generate the same output frequency. For instance, a multi-sinusoidal
input with frequencies 1, 2, 3, 4 (in Hertz) would generate many frequency
overlaps such as $1=2-1$ or $2+3=1+4$ and so on. In presence of
frequency overlaps, it is not possible to unambiguously measure the
nonlinear frequency interactions. In the previous example, the measurement
at 5 Hz is ambiguous because one is unable to distinguish between
the contributions of $2+3$ and $1+4$. In order to avoid this problem,
the QSA was used with a flexible algorithm generating incommensurable
frequencies. This approach is based on a practical measurement technique,
namely harmonic probing on Volterra kernels (\citealp{Boyd1983};
\citealp{Victor1980}).

Since harmonic and intermodulation responses also exist for nonlinearities
of higher degrees, for example third order intermodulation products
$f_{1}+f_{2}+f_{3}$, it is important to ensure that the neurons mainly
manifest quadratic nonlinearities, otherwise the results would be
significantly contaminated. For this, a necessary, but not sufficient
condition, consists of using relatively small stimulus amplitudes
in such a way that only linear and quadratic responses are significant.
This approach was used in our previous piecewise linear analysis (\citealp{Murphey1995}),
hence the term piecewise quadratic analysis can be used to describe
the QSA extension. The influence of the input amplitude on the harmonic
response has been investigated previously (see \citealp{Moore1980};
\citealp{FitzHugh1983}). In order to ensure that the stimulus amplitudes
were sufficient to overcome spontaneous noise and avoid significant
higher order responses, several algorithms described elsewhere (\citealp{Magnani2010})
have been implemented in MATLAB to verify that the experimental traces
are time invariant for both linear and quadratic outputs, and that
the signal can be adequately reconstructed by quadratic analysis.

It will be shown that the oscillatory type D neurons of the PHN have
quadratic responses over a range of subthreshold membrane potentials,
namely they convert limited amplitude and bandwidth input signals
to wider bandwidth and more complex output responses as mentioned
above for nonlinear responses. Under normal physiological conditions
of current clamp, a depolarization activates gNaP conductance that
in turn increases the impedance and consequently reduces the electrotonic
length. Type B neurons show similar effects and in addition, they
have a significant increase in the electrotonic length at hyperpolarized
membrane potentials because of the activation of a hyperpolarization
activated conductance (\citealp{Erchova2004}).

Voltage clamp experiments were done to partially control the oscillatory
and bistable responses of PHN neurons in order to analyze the nonlinear
membrane properties of both the somatic and dendritic regions of these
neurons. Indeed, an important advantage of a quantitative voltage
clamp analysis of central neurons is the exploitation of the space
clamp problem as a way to separate somatic and dendritic responses,
namely the voltage clamp current measured from the voltage controlled
soma is generally dominated by the unclamped voltage responses of
the dendrites (\citealp{Moore1999}). Thus, this current can be used
as a measure of both linear and nonlinear dendritic potential responses
while the somatic membrane potential is voltage clamped. These additional
currents can flow because of a potential difference between the soma
and the rest of the dendrite. In fact, these additional currents reflect
the behavior of the dendritic membrane potential that can be taken
into account with multi-compartmental models.

Previous voltage clamp measurements have been done using signals of
small amplitude to obtain steady state linear responses at different
membrane potentials. These measurements were done over a potential
range to obtain a quantitative description of the voltage dependent
conductances and have allowed the construction of neuronal multi-compartmental
models of both type B and D PHN neurons (\citealp{Idoux2008}). These
models allow an estimation of both somatic and dendritic membrane
properties from somatic voltage clamp experiments that probe the soma
and all regions of the dendritic structure.

When the stimulus amplitudes are sufficiently small to elicit linear
responses, both the voltage clamp and the current clamp generate equivalent
linear results, however this is not the case for nonlinear responses.
The present series of voltage clamp experiments is a quadratic extension
of the steady state linear analysis. Voltage clamped neurons show
two kinds of nonlinearities : first, space clamped somatic ionic currents,
and second, ionic currents in the soma due to an unclamped dendritic
membrane. However, in a similar current clamp experiment, the nonlinear
behaviors measured from the soma are caused by the voltage responses
of both the somatic and the dendritic membranes. Due to this asymmetry
between voltage clamp and current clamp, there is no obvious way to
predict the voltage clamp nonlinear response from the current clamp
nonlinear response nor the converse. It is well known in the linear
case that the admittance from voltage clamp, $Y\left(f\right)=\frac{\hat{I}\left(f\right)}{\hat{V}\left(f\right)}$,
is the inverse of the impedance from current clamp $Z\left(f\right)=\frac{\hat{V}\left(f\right)}{\hat{I}\left(f\right)}$
, that is $Y=Z^{-1}$ where $\hat{I}$ or $\hat{V}$ refers to Fourier
transforms of I and V, respectively. In the nonlinear case, with nonoverlapping
frequencies, the quadratic response for the voltage clamp is defined
by QSA as $B_{vc}\left(f_{1},f_{2}\right)=\gamma_{f_{1},f_{2}}\frac{\hat{I}\left(f_{1}+f_{2}\right)}{\hat{V}\left(f_{1}\right)\hat{V}\left(f_{2}\right)}$
and for the current clamp by $B_{cc}\left(f_{1},f_{2}\right)=\gamma_{f_{1},f_{2}}\frac{\hat{V}\left(f_{1}+f_{2}\right)}{\hat{I}\left(f_{1}\right)\hat{I}\left(f_{2}\right)}$,
where $\gamma_{f_{1},f_{2}}$ is a symmetry factor (\citealp{Magnani2010}).
In general $B_{vc}\left(f_{1},f_{2}\right)$ and $B_{cc}\left(f_{1},f_{2}\right)$
are not reciprocally equivalent because of the asymmetry discussed
above. Hence, this is an important conceptual difference between linear
and nonlinear analysis, which also plays a role in interpretation
of current versus voltage clamp experiments.

The terms, $\hat{V}\left(f_{1}+f_{2}\right)$ and $\hat{I}\left(f_{1}+f_{2}\right)$
where $f_{1}$ and $f_{2}$ are either positive or negative frequencies
refer to all complex values of $\hat{V}$ and $\hat{I}$ at the interactive
and harmonic quadratic frequencies. The quadratic response function
can then be represented as magnitude and phase plots versus $f_{1}$
and $f_{2}$.

\subsubsection*{Matrix reduction}

Since $B_{vc}\left(f_{1},f_{2}\right)$ is difficult to interpret,
it is convenient to reduce it to a diagonal matrix through eigendecomposition
methods. Similar methods have been used in quantitative neuronal analyses,
for example singular value decomposition (\citealp{Lewis2002c}) or
principal component analysis (\citealp{Haas2007}). To make this reduction,
it is important to note that the matrix $Q_{vc}$ obtained by row
flipping $Q_{vc}\left(f_{1},f_{2}\right)=B_{vc}\left(-f_{1},f_{2}\right)$
is actually Hermitian. As a consequence, $Q_{vc}$ can be reduced
to a diagonal matrix $D$ such that $Q_{vc}=P^{*}DP$ where $P$ is
a unitary matrix and $P^{*}$ its complex conjugate transpose. The
unitary matrix $P$ contains no information about the magnitude and
can be viewed as a kind of generalization of the phase. The magnitude
is entirely encoded in the diagonal matrix $D$. The elements of $D$
are called eigenvalues. Each column of the matrix $P^{*}$ is a special
vector called an eigenvector, whose coordinates are expressed relatively
to the stimulus frequencies. The amplitude of each eigenvalue indicates
the relative contribution of the corresponding eigenvector to the
quadratic response. Practically, the eigendecomposition can be interpreted
as a reduction of the quadratic neuronal function to a set of quadratic
filters in which eigenvalues play the role of amplitudes.

\begin{figure}
\includegraphics[width=\columnwidth,bb=0 0 432 351]{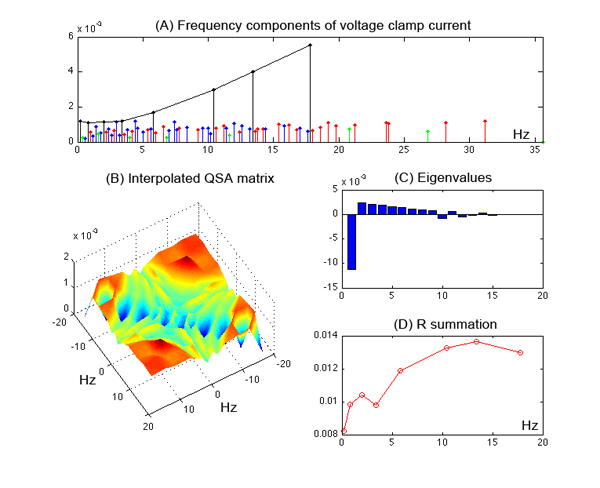}
\caption{\scriptsize{Voltage clamp measurement at -50 mV on a type D neuron.\protect \\
(A) Amplitude of the evoked linear current responses (black points
with interpolation line) and nonlinear frequency components (colored
points). The stimulation show an increasing amplitude as the frequency
increases. The nonlinear interactions are shown as individual points
for $f_{i}+f_{j}$ in red, $\left|f_{i}-f_{j}\right|$ in blue, $2f_{i}$
in green where $f_{i}$ and $f_{j}$ are positive.\protect \\
(B) Three dimensional plot of the magnitudes of the interpolated
QSA matrix for a continuous range of frequencies. Each axis is indexed
by 8 negative followed by 8 positive frequencies, which are shown
as a 16x16 square matrix.\protect \\
(C) Eigenvalues of the QSA matrix, illustrating the dominance of
one eigenvalue. The abscissa is shown with numbers indicating each
of the 2N eigenvalues by decreasing magnitudes.\protect \\
(D) Summation $R\left(j\right)$ showing the influence of each
stimulation frequency in the quadratic responses.\protect \\
The voltage command was a multsinusoidal stimulation, which had
a standard deviation of 2.85 mV. For all plots, the frequency components
were computed with the MATLAB command FFT divided by the number of
points. The current was measured in nA and the voltage in mV. The
stimulation was constructed from the nonoverlapping set (0.2, 0.8,
2, 3.4, 5.8, 10.4, 13.4, 17.8) in Hz.} \label{fig:nfig1a}}

\end{figure}

The responses at the stimulus frequencies are shown in Figure \ref{fig:nfig1a}A
as interpolated black points. The color coded points illustrate the
smaller amplitude current responses at the quadratic frequencies.
At low frequencies both the linear and nonlinear responses are quite
comparable despite the fact that the standard deviation of the imposed
voltage clamp stimulus was only 2.85 mV. In Figure \ref{fig:nfig1a}B,
each cell $\left(i,j\right)$ of the QSA matrix encodes the magnitude
of the corresponding quadratic interaction $\left(f_{i},f_{j}\right)$.
Informally, the QSA matrix can be viewed as a quadratic generalization
of the admittance, thus it is an intrinsic characterization of the
measured neuronal response. The interpolations were performed by the
MATLAB command GRIDDATA (linear method) in order to represent the
responses in 3D color plots over a continuous range of frequencies.
As a characteristic function, the QSA matrix is independent of the
stimulus amplitude, however the current responses become proportionally
insignificant as the stimulus approaches zero. The responses of higher
orders are insignificant for stimulation amplitudes below \textpm{}3
mV.

The QSA matrix, $Q_{vc}$, is the core mathematical object which encodes
the total effect of the pairwise interactions at the interactive frequencies.
Each matrix cell is located at the intersection of two input frequencies.
In this way, the one dimensional admittance function $Y\left(f\right)$
is generalized to a two dimensional quadratic function $Q_{vc}\left(f_{1},f_{2}\right)$.
The QSA along with linear analysis can be used to reconstruct the
signal using $Y$ and $Q_{vc}$. Figure \ref{fig:nfig1a}B shows that
the maximum amplitude for the interactive frequencies occurs at the
intersection of 2 Hz and 10.4 Hz ($f_{1}+f_{2}$ ).

The eigendecomposition of these data strongly suggests that depolarized
type D neurons are dominated by a single eigenvalue as illustrated
in Figure \ref{fig:nfig1a}C. Even when two eigenvalues are required
to adequately describe the response, the eigendecomposition provides
a remarkably compact representation of the nonlinear response that
otherwise can only be quantitatively described by complex differential
equations involving large numbers of dendritic compartments. It would
appear that the neuronal function becomes more complex, in the sense
of information processing, as the number of significant eigenvalues
increases.

\subsubsection*{Matrix summation}

The previous matrix reduction has the great advantage of being reversible,
such that from $P$ and $D$ one can exactly recover $Q_{vc}$. There
exists a coarser simplification by summing the QSA matrix by columns
to obtain a vector indexed by the stimulation frequencies. The summation
is defined as follows :

\[
R\left(j\right)=\sum_{i}\left|Q_{vc}\left(i,j\right)\right|\]
The values of R are illustrated in Figure \ref{fig:nfig1a}D. The
advantage of the R functions is that they can be presented as classical
Bode plots. Moreover, each R function can be intuitively interpreted
as a measure of the influence of each individual stimulation frequency
on the nonlinear responses involving its interaction with all other
stimulation frequencies. Hence, matrix summation is especially well
suited to superimpose and compare with the piecewise quadratic analysis
for different steady state responses.

\begin{figure}
\includegraphics[width=\columnwidth,bb=0 0 432 308]{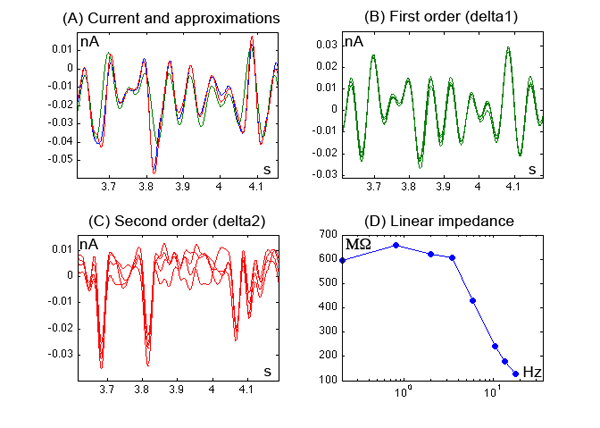}

\caption{\scriptsize{Nonlinear and linear responses of the voltage clamped type D neuron
of Figure \ref{fig:nfig1a} .\protect \\
(A) Superposition of the averaged data with the first (green) and
second order (red) responses to describe the total neuronal response
(blue).\protect \\
(B) Superposition of only first order responses for four traces.\protect \\
(C) Superposition of only second order responses for four traces
which provides an indication of the synchrony and reproducibility
of the second order components in the data for significantly large
amplitudes.\protect \\
(D) Linear impedance interpolated from the stimulating frequencies.}
\label{fig:methods1a}}

\end{figure}

Figure \ref{fig:methods1a}A shows a reconstruction of the measured
currents with the first and second order responses. The contribution
of the nonlinear frequencies can be quantitatively expressed by comparing
the ratios of the spectral energy of the linear versus linear + quadratic
responses to the total spectral energy for the entire range of frequencies.
Since the maximum stimulus frequency was 17.8 Hz, a frequency range
of 36 Hz > 2 x 17.8 was selected for the total sum. Clearly, the quadratic
reconstruction is much more accurate than the linear one, which is
confirmed by evaluating the signal energy, 61\% for the linear analysis
against 96\% for the quadratic analysis. The discrepancy between the
second order ratio and 100\% shows how well a second order approximation
fits the total response, as well as an indication of the presence
of other higher order responses for the frequency range selected.
Higher order responses clearly are more prominent with a depolarization
due to the augmentation of the nonlinear behavior of both second and
higher order responses.

Since the second order responses have relatively low amplitudes it
is essential to ensure that the contributions of other noise sources,
such as synaptic events or membrane ion channel fluctuations, do not
significantly contribute to the observed responses at the harmonic
and interacting frequencies. The synchrony and reproducibility of
the data is shown by superimposing first order (Figure \ref{fig:methods1a}B,
delta1) and second order (Figure \ref{fig:methods1a}C, delta2) computed
responses for four sequential measurements. A time invariance correlation
function was used to determine an optimal stimulus amplitude, which
is large enough to overcome the spontaneous noise of the neuron and
not too great to evoke significant higher order responses (\citealp{Magnani2010}).
Finally, the usual linear impedance, $Z\left(f\right)=\frac{\hat{V}\left(f\right)}{\hat{I}\left(f\right)}$,
is shown in Figure \ref{fig:methods1a}D.

\section*{Results}

\subsection*{Type D neurons}

It is difficult to get accurate measurements of the quadratic responses
in current clamp for the type D neurons, mainly due to uncontrolled
spontaneous oscillations. Thus, voltage clamp experiments were done
to control the oscillations and measure the negative current associated
with the persistent sodium conductance gNaP. The nonlinear responses
evoked by 5-10 mV voltage clamp stimuli can also be blocked (not shown)
by riluzole as described in an earlier steady state piecewise linear
analysis of gNaP (\citealp{Idoux2008}).

\begin{figure}
\includegraphics[width=\columnwidth,bb=0 0 432 382]{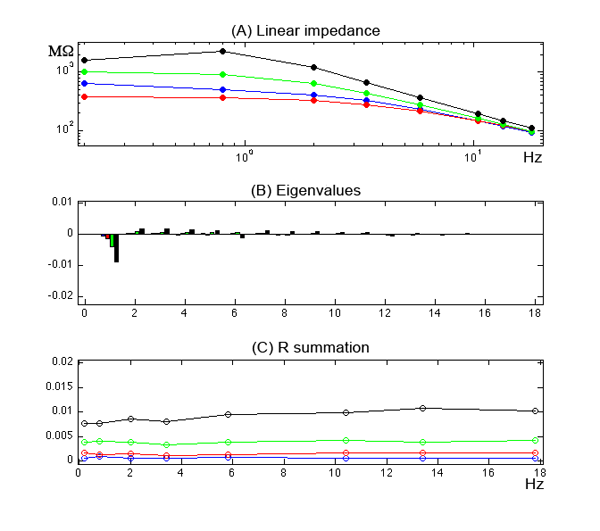}
\caption{\scriptsize{Voltage clamp experiments on a PHN type D neuron show the effect of
an activated gNaP by comparing both the linear and quadratic voltage
clamp responses at -70, -60, -55, -50 mV (blue, red, green, and black
respectively).\protect \\
(A) Linear impedance, however it should be noted that the stimulus
is a controlled voltage that leads to a current response. The depolarization
markedly increases the impedance.\protect \\
(B) Eigenvalues of the QSA matrix. The abscissa is shown with numbers
indicating each of the 2N eigenvalues by decreasing magnitudes.\protect \\
(C) Magnitude of R summations of the corresponding QSA matrix\emph{.}
They also increase during a depolarization.\protect \\
Statistically in five neurons the maximum eigenvalue increased
for a change of membrane potential of -60 to -50 mV from 0.002\textpm{}0.001
to 0.010\textpm{}0.001 with p-value=0.00004. The mean value of R\textpm{}SD
increased from 0.005\textpm{}0.003 to 0.023\textpm{}0.004 with p-value=0.0003.}
\label{fig:Dneuron}}

\end{figure}

Figure \ref{fig:Dneuron}A shows the increased linear impedance magnitude
evoked by progressive depolarizations. Since gNaP can be seen as a
negative conductance, its activation by depolarized membrane potentials
reduces the total conductance of the cell, which leads to an increase
of impedance. The nonlinear responses are indicated by eigenvalues
(Figure \ref{fig:Dneuron}B) and R summation (Figure \ref{fig:Dneuron}C),
whose magnitudes clearly increase with depolarization. For this neuron,
the spectral energy analysis led to ratios 99\% and 100\% at -60 mV,
and 77\% and 98\% at -50 mV for the linear versus linear + quadratic
responses, respectively.

\subsection*{Type B neurons}

\begin{figure}
\includegraphics[width=\columnwidth,bb=0 0 432 417]{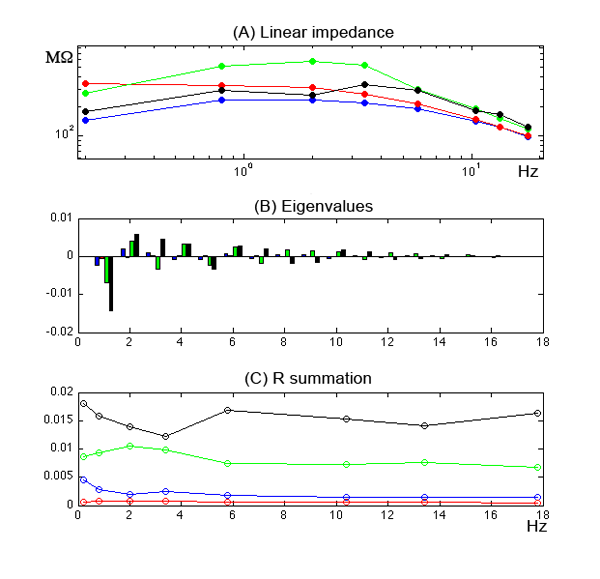}
\caption{\scriptsize{Voltage clamped type B neuron at -75, -60, -46, -41 mV (blue, red,
green, black, respectively).\protect \\
(A) Linear impedance showing resonance induced by the depolarization.\protect \\
(B) Multiple eigenvalues increasing with depolarization in contrast
to a single dominant eigenvalue observed for type D neurons. The abscissa
numerically labels the eigenvalues by decreasing magnitude.\protect \\
(C) Magnitude of R summations.\protect \\
These plots show that nonlinear responses are present at both hyperpolarized
and depolarized membrane potentials.\protect \\
Statistically in six neurons, for a positive membrane potential
change of 5 mV in the range -60 to -40 mV, the maximum value of the
eigenvalue increased from 0.004\textpm{}0.003 to 0.014\textpm{}0.009
with p-value=0.008 and the mean value of R\textpm{}SD increased from
0.009\textpm{}0.006 to 0.03\textpm{}0.02 with p-value=0.01.}\label{fig:Bneuron}}

\end{figure}

It has been shown previously that type B neurons have a prominent
gNaP, which often leads to net inward current carried by Na$^{+}$
for a limited range of voltage clamped depolarized membrane potentials
(\citealp{Idoux2008}). Voltage clamped data in Figure \ref{fig:Bneuron}
illustrate that type B nonlinear responses are significantly enhanced
by depolarization and in addition, often show a resonant enhancement
of the impedance. The impedance shows a maximum at an intermediate
depolarization and a shift to a higher resonance frequency with further
depolarization. In contrast, Figures \ref{fig:Bneuron}B and C show
that both the eigenvalues and the R summation values increase monotonically
with depolarized membrane potentials. Interestingly, the number of
significant eigenvalues required to describe the nonlinear response
is generally two or more, unlike the single eigenvalue usually needed
for type D neurons. For this neuron, the spectral energy analysis led to ratios 99\% and 100\% at -60 mV, and 82\% and 96\% at -41 mV for the linear versus linear + quadratic responses, respectively.

\subsection*{Effect of gNaP and NMDA activation}

\begin{figure}
\includegraphics[width=\columnwidth,bb=0 0 432 459]{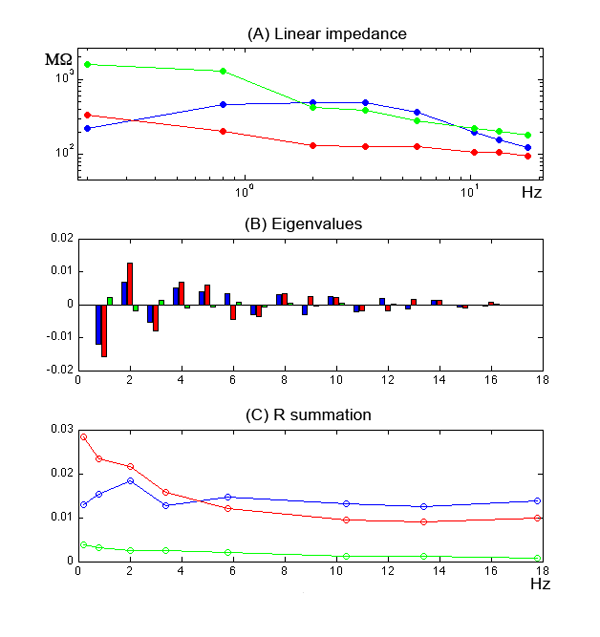}
\caption{\scriptsize{Effect of NMDA (red) and NMDA+TTX (green) compared to control (blue)
on a voltage clamped type D neuron at -40 mV.\protect \\
(A) Effect of 25 $\mu$M NMDA on linear impedance. The impedance
decreases with NMDA (red) and increases with 2 $\mu$M TTX
(green) compared to the control (blue)\protect \\
(B) The NMDA plots (red) illustrate that the nonlinear responses
are increased by the presence of NMDA. Moreover, TTX (green), which
blocks\emph{ }gNaP, when added to NMDA shows decreased eigenvalues\emph{.}
The abscissa numerically labels the eigenvalues by decreasing magnitude.\protect \\
(C) Effect of NMDA and NMDA+TTX on R summation. The R values are
in inverse order to the impedance magnitudes.\protect \\
Statistically, pooled data from five neurons (three of type B and
two of type D) in the presence versus absence of 25-50 $\mu$M
NMDA over the range -66 to -40 mV, showed a significant difference
in the mean values of R summation (p-value = 0.026). Similarly the
maximum eigenvalues were significantly different with a p-value of
0.035. TTX reduces the nonlinear effects of NMDA to control values
or less (p-value = 0.09). An insignificant p-value of 0.2 was found
comparing control with NMDA + TTX treated neurons.}\label{fig:NMDAexperiment}}

\end{figure}

Nonlinear responses are likely to be enhanced by the activation of
dendritic NMDA receptors, which would occur during synaptic activity
of the neural integrator network. NMDA activation could trigger dendritic
bistable responses that contributes to maintaining a particular firing
rate after an input impulse. Figures \ref{fig:NMDAexperiment}B and
C show that the addition of NMDA clearly enhances nonlinear response
of a type D neuron consistent with the trigger hypothesis. The mean
value of R and the maximum eigenvalue for all neuronal types increased
significantly with the addition of NMDA. TTX reduces the nonlinear
effects of NMDA to control values or less, however it is still capable
of inducing potential oscillations in current clamp (not shown). Control (with gNaP, without NMDA) versus NMDA + TTX treated neurons (without gNaP, with NMDA) did not show a significant p-value. Thus, NMDA and gNaP are synergistic in their action,
where the combined effects are generally greater than either alone.
Since the normal physiological activation of NMDA receptors is due
to transient synaptic currents, the non-inactivation of gNaP and membrane
potential bistability contribute to the maintenance of a depolarized
potential. In conclusion gNaP appears to be essential for sustained
nonlinear effects induced by NMDA activation and thus would be critical
for the operation of the neural integrator.

\subsection*{Type D and B neuronal model simulations}

In order to provide an interpretation of these experimental results,
numerical simulations have been done using previously published models
of types D and B for rat PHN neurons (\citealp{Idoux2008}). These
models have been constructed from a piecewise linear analysis to fit
parameters of nonlinear differential equations in voltage clamp. Unless
otherwise indicated, both models had three uniformly distributed voltage
dependent conductances, gK, gNaP and gH with a soma and eight dendritic
compartments. Typically the gH of type D neurons is quite small and
could be neglected.

\begin{figure}
\includegraphics[width=\columnwidth,bb=0 0 432 354]{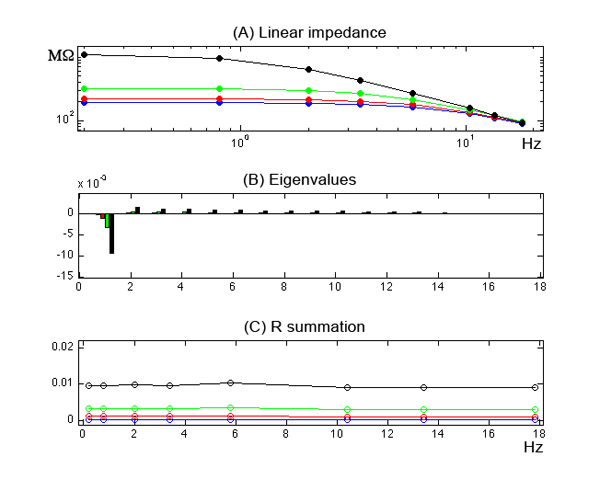}
\caption{\scriptsize{Model of voltage clamped type D PHN neuron. Simulations were done
at -70, -60, -55, -50 mV (blue, red, green, black, respectively).\protect \\
(A) Linear impedance. The depolarization markedly increases the
linear impedance.\protect \\
(B) Eigenvalues of the QSA matrix. The abscissa numerically labels
the eigenvalues by decreasing magnitude.\protect \\
(C) Magnitude of R summations of the corresponding QSA matrix\emph{.}\protect \\
The following parameter values were used: csoma = 20.5 pF; aratio
= 3.77; elength = 0.54; gleak = 1.2 nS(1.37 nS); vleak = -53 mV; gk
= 1.18 nS; vn = -32 mV; sn = 0.06(mV)$^{-1}$ (0.05 (mV)$^{-1}$);
tn = 10 msec; vk = -87 mV; gH = 0; sm = 0.06 (mV)$^{-1}$; tm = 150
$\mu$sec; vNa = 77 mV; vm = -38 mV (-35 mV); gNaP= 0.9 nS
(0.6 nS). Parameter values shown in parentheses are default values
defined in \citealp{Idoux2008}.}\label{fig:Dmodel}}

\end{figure}

Simulations done with the published average parameter values were
consistent with the voltage clamp data, however some appropriate
modifications of the average parameters were done for the comparison
with the data for the individual neurons of Figures \ref{fig:Dneuron}
and \ref{fig:Bneuron}. Four potentials are shown to cover the range
observed in the experiments. The type D data and model generally show
a dominant eigenvalue (see -50 mV in Table \ref{tab:table1} and Figure
\ref{fig:Dmodel}).

\begin{figure}
\includegraphics[width=\columnwidth,bb=0 0 432 451]{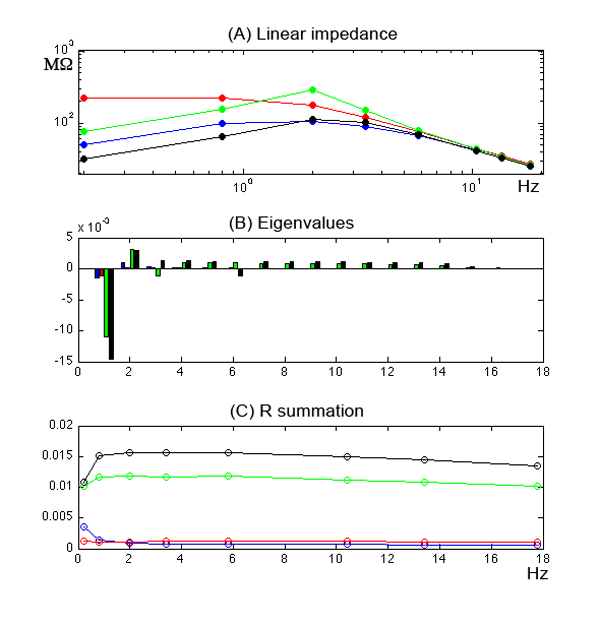}
\caption{\scriptsize{Model of voltage clamped type B neuron. Simulations were done at -76,
-60, -46, -41 mV (blue, red, green, black, respectively).\protect \\
(A) Linear impedance. The magnitude shows a resonant maximum at
-46 mV.\protect \\
(B) Eigenvalues of the QSA matrix. The abscissa numerically labels
the eigenvalues by decreasing magnitude.\protect \\
(C) Magnitude of R summations of the corresponding QSA matrix\emph{.}\protect \\
The parameter values were as follows: csoma = 0.0000995 $\mu$F
(0.0000265); aratio = 2.6 (2.85); elength = 0.31 (0.37); gleak = 0.001297
$\mu$S (0.000878); vleak = -56.61 $\mu$S (-54.9);
gk = 0.005555 $\mu$S (0.0024); vn = -38.7 mV (-35); sn = 0.07715/mV
(0.05); tn = 0.66 sec (.09); vk = -87 mV; gH = 0.00644 $\mu$S
(0.00591); vq = -79.7 mV (-63.9); sq = -0.0667/mV (-0.0647); tq =
1.75 sec (8.43); vk2 = -43 mV; gNaP = 0.00147 $\mu$S (0.00131);
vm = -35 mV; sm = 0.05/mV (.056); tm = 0.00015 sec; vnmda = 77 mV;
Parameter values shown in parentheses are default values from \citealp{Idoux2008}.}\label{fig:Bmodel}}
\end{figure}

The simulations of the type B model in Figure \ref{fig:Bmodel} show
behavior similar to the data of Figure \ref{fig:Bneuron}, showing
two or more significant eigenvalues compared to an essentially single
dominant eigenvalue for the type D model (Figure \ref{fig:Dmodel}).
However Tables \ref{tab:table1}  and \ref{tab:table2} indicate
that the number of significant eigenvalues for both neuronal types
is dependent on both the membrane potential and specific parameter
values. For example, at -40 mV with or without NMDA, Figure \ref{fig:NMDAexperiment}
B shows a type D neuron with multiple significant eigenvalues (also,
see Table \ref{tab:table1}).

In addition, the monotonic increase in the R summation of the type
B model nonlinear responses with depolarization contrasts with the
peaking of linear impedance increase similar to that observed in the
type B neuron of Figure \ref{fig:Bneuron}. In general the impedance
increase with depolarization is caused by the activation of the gNaP
negative conductance that balances the other positive conductances.
An impedance maximum, often with a resonance, occurs because the impedance
decreases with further depolarization due to an increased gK.

The nonlinear responses of type B model simulations, in contrast to
type D, show that the quadratic response is greatly reduced in magnitude
at -60 mV, in part because the type B model has a greater density
of voltage dependent gNaP conductances on a relatively more compact
dendritic structure than found for type D neurons. In addition, the
nonlinear responses in the type B model, as in the data of Figure
\ref{fig:Bneuron}, are enhanced at hyperpolarized potentials directly
due to the gH conductance.

\begin{figure}
\includegraphics[width=\columnwidth,bb=0 0 432 373]{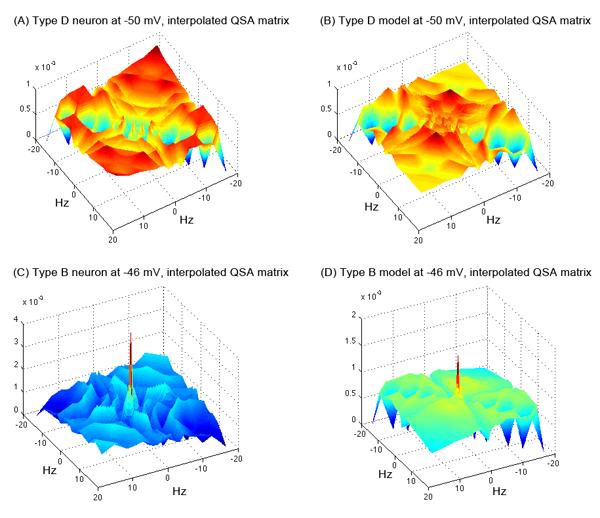}
\caption{\scriptsize{Interpolated QSA plots of type D and B neurons. (A) type D neuron
of Figure \ref{fig:Dneuron} at -50 mV. (B) Type D model of Figure
\ref{fig:Dmodel} at -50 mV. (C) Type B neuron of Figure \ref{fig:Bneuron}
at -46 mV. (D) Type B model of Figure \ref{fig:Bmodel} at -46 mV.}\label{fig:QSA-plots}}
\end{figure}

The use of nonoverlapping frequencies is required for the construction
of QSA matrices from experimental measurements. Nevertheless, it is
possible to do a coarse interpolation for other  frequencies as plotted
in Figure 8. The interpolated QSA matrices for type D and B neurons
show striking differences, which also occur in their mathematical
models. Figures \ref{fig:QSA-plots}A and \ref{fig:nfig1a}B show
local peaks in the QSA plots at two membrane potentials for the same
type D neuron, which are comparable to the type D model simulation
in Figure \ref{fig:QSA-plots}B. By contrast the type B neuron and
model (Figure \ref{fig:QSA-plots}C, D) show a prominent low frequency
peak distinctly different than observed in type D neurons. The QSA
matrix clearly provides significantly more information about the neuronal
behavior than linear parameters and has the possiblilty of being extended
to all frequencies of interest.

In order to test the hypothesis that the dendritic conductances are
mainly responsible for the nonlinear responses, simulations were done
with gNaP reduced in the soma and unchanged in the dendrite or the
converse as indicated in Tables \ref{tab:table1} and \ref{tab:table2}.
In type D neurons, decreasing the soma gNaP has a lesser effect (79\%)
on the R summation compared with decreasing the dendritic gNaP (13\%).
Simulations of type B model neurons showed a lesser effect for a gNap
reduction in the more compact dendrite (34\%). Such simulations support
the hypothesis that the nonlinear responses measured under voltage
clamp conditions are dominated by dendritic responses, while the linear
responses are determined by both the soma and dendrite. Since the
somatic region has good voltage space clamp control, one would expect
a dominant linear current response from the somatic conductances at
all membrane potentials, however the significantly larger dendritic
membrane area is not space clamped leading to uncontrolled voltage
excursions and greater nonlinearites. In conclusion, these simulations
suggest that nonlinear responses in voltage clamped neurons are dominated
by active dendritic structures, when their electrotonic lengths are
above 0.3.

\begin{table}
\begin{tabular}{|c|c|c|c|c|c|c|}
\hline 
soma-Na & dendrite-Na & Vm & R{\scriptsize mean} & $\mu$1 & $\mu$2 & $\mu$2/$\mu$1(\%)\tabularnewline
\hline
\hline 
1 & 1 & -40 & 0.0162 & +0.0160 & -0.0049 & 31\tabularnewline
\hline 
1 & 1 & -50 & 0.0095 & -0.0095 & +0.0015 & 16\tabularnewline
\hline 
0 & 0 & -50 & 0.0011 & +0.0011 & -0.0002 &  18\tabularnewline
\hline 
0.01 & 1 & -50 & 0.0075 & -0.0074 & +0.0012 & 16\tabularnewline
\hline 
1 & 0.01 & -50 & 0.0012 & -0.0012 & +0.0003 &  25\tabularnewline
\hline
\end{tabular}
\caption{\scriptsize{Type D model simulations. The voltage clamped somatic membrane potential
is given along with presence or absence of conductances indicated
by 1 or 0. The value of gNaP in either the soma or all dendritic compartments
was reduced to 1\% indicated by 0.01 in the labeled columns. The mean
values of R summation and the two largest eigenvalues $\mu$1
and $\mu$2 are given for each simulation. The last column
is $\mu$2 represented as the percentage of the absolute $\mu$1
value. A positive $\mu$1 occurs when gK has a greater effect
than gNaP.}\label{tab:table1}}

\end{table}

\begin{table}
\begin{tabular}{|c|c|c|c|c|c|c|}
\hline 
soma-Na & dendrite-Na & Vm & R{\scriptsize mean} & $\mu$1 & $\mu$2 & $\mu$2/$\mu$1(\%)\tabularnewline
\hline 
1 & 1 & -40 & 0.0140 & -0.0141 & +0.0027 & 19\tabularnewline
\hline 
1 & 1 & -50 & 0.00652 & -0.0063 & +0.0018 & 29\tabularnewline
\hline 
0 & 0 & -50 & 0.00104 & +0.0018 & -0.0008 & 44\tabularnewline
\hline 
0.01 & 1 & -50 & 0.00467 & -0.0044 & +0.0020 & 45\tabularnewline
\hline 
1 & 0.01 & -50 & 0.00221 & -0.0019 & +0.0015 & 79\tabularnewline
\hline
\end{tabular}

\caption{\scriptsize{Type B model simulations as in Table \ref{tab:table1}. The mean values
of R summation and the two largest eigenvalues $\mu$1 and
$\mu$2 are given for each simulation. When gNaP in the dendrite
is reduced, the negative $\mu$1 decreases and becomes as large
as the unchanged positive $\mu$2 (79\%), which suggests structurally
different quadratic filters for type B compared to type D models.}
\label{tab:table2}}

\end{table}

\section*{Discussion}

The neurons of the prepositus hypoglossi nucleus (PHN) provide a useful
system to investigate nonlinear behavior, such as persistent activity
to maintain eye position. The oscillatory character of some of these
neurons is similar to that observed in the stellate neurons in layer
IV of entorhinal cortex (\citealp{Haas2002}; \citealp{Schreiber2004}),
which are also involved in the processing of orientational information.
PHN neurons are part of the brainstem that receives head velocity
signals and integrates them to control eye position for the stabilization
of an image at the center of the visual field during head rotation.
This specific processing is called neural integration (\citealp{Aksay2007})
due to an analogy with integration in mathematical calculus.

In order to understand how the PHN neural network can perform neural
integration, it is important to understand the biophysical properties
of individual neurons involved in the circuitry. Single neurons of
the PHN show oscillatory and bistable nonlinear properties that are
likely to be involved in the operation of the neural integrator. Since
models based on recurrent excitation, even including lateral inhibition,
are not sufficiently robust, a number of theoretical papers have suggested
that the nonlinear properties of individual neurons are essential
for the network behavior of the neural integrator (\citealp{Koulakov2002};
\citealp{Goldman2003}). Thus, the finding that oscillatory nonlinear
behavior is clearly present in the neurons involved in the eye movement
circuitry (\citealp{Idoux2006}) lends strong support to these theoretical
notions. Dendritic characteristics (\citealp{Johnston2008}) are potentially
critical for the function of these neurons, which in addition have
also shown to be different for the two main classes of PHN neurons (\citealp{Idoux2008}).

Thus, type D oscillatory neurons have bistability properties, which
are consistent with neural integrator models that rely on remote dendritic
processing (\citealp{Goldman2003}; \citealp{Idoux2008}). The experiments
and analysis in this paper strongly support the hypothesis that type
D neurons have persistent sodium channels in their dendrites, which
would promote remote bistable potentials shifts leading to persistent
activity. Type B neurons have an even higher density of persistent
sodium channels and their passive electrotonic length is less than
type D. As a consequence, the type B neurons at moderate depolarizations
would have a more uniform potential throughout the less isolated dendritic
tree.

The gNaP conductance in the dendrites can easily be activated by the
synaptic stimulation of the NMDA receptors likely to occur during
normal physiological activity. In addition, NMDA activation enhances
total nonlinearities, which are shown for a type D neuron and also
observed in type B neurons (see Figure \ref{fig:NMDAexperiment}).
In current clamp conditions, both types of neurons show marked potential
oscillations in the presence of 25-50 $\mu$M NMDA (\citealp{Idoux2006}).
Clearly the level of NMDA activation expected during synaptic activation
would be much less, however it is likely to be sufficient to evoke
significant nonlinear responses due to the gNaP in both type B and
D neurons. In this regard, the neural integrator model of \citeauthor{Koulakov2002}
(2002), depends on NMDA synapses that activate gNaP dependent bistable
states in dendritic compartments.

The quantitative measurement of the biophysical properties of intact
neurons is seriously compromised by the inherent inability to voltage
clamp the electrotonic structure of the dendritic tree. In general
whole cell measurements are restricted to patch clamp electrodes placed
in the soma, which makes it difficult to infer the remote properties
of the dendrites. Previous piecewise linear analyses have permitted
the development of realistic neuronal models, however it has been
difficult to separate the properties of the dendrites from the soma
unless patch clamp electrodes can be placed in the dendrites. The
new approach described in this paper takes advantage of the space
clamp problem associated with voltage clamping neurons. The quadratic
analysis has shown that it is possible to characterize the nonlinear
behavior of the uncontrolled dendritic membrane voltage responses
while maintaining voltage clamp control of the somatic membrane. When
the dendritic electrotonic structure is remote with relatively large
surface areas compared to the soma, the nonlinear behavior of the
dendritic membrane dominates the soma. This is especially the case
in a voltage clamp experiment because of the lack of dendritic potential
control.

It was found that responses in the range of \textpm{}(5-10) mV could
be well described by quadratic nonlinearities suggesting that nonlinearities
of higher degrees only add marginal improvement. Thus, the quadratic
response is likely to sufficiently capture most of the nonlinear behavior
of neuronal systems except for extremely large synaptic inputs. The
quadratic functions are quite sensitive to the mean membrane potential
and appear to be valid for a range of sinusoidal inputs. This behavior
extends significantly the validity of the quantitative quadratic description.

The quadratic functions are computed on particular sets of nonoverlapping
frequencies, which can be interpolated over a continuous range of
frequencies as illustrated in Figures \ref{fig:nfig1a}B and \ref{fig:QSA-plots}.
Thus, they provide a significantly concise description of the neuronal
behavior and could potentially be used as computational devices that
would be independent of nonlinear differential equations. Practically,
this could be an alternative approach to large scale neural network
simulations.

The estimation of the parameters of both the voltage dependent conductances
and the electrotonic structure have shown quantitative differences
between type B and D neurons (\citealp{Idoux2008}). In addition,
the nonlinear analysis in this paper suggests that the number of significant
eigenvalues is greater for the type B versus the type D neurons and
their individual models. Thus, the measured nonlinearities seems to
be structurally different between type B and D, namely the two corresponding
types of quadratic functions are intrinsically different (Figure \ref{fig:QSA-plots}).
In general the dominant eigenvalue was negative, which is related
to the negative slope conductance due to gNaP. Type D model simulations
at large depolarizations (see Table \ref{tab:table1} at -40 mV) show
that the maximum eigenvalue ($\mu$1) is positive, consistent
with a positive slope conductance due to an increased outward potassium
current. In contrast, the greater gNaP of the type B model maintains
a negative $\mu$1 at -40 mV (Table \ref{tab:table2}). Both
type D and B models show positive $\mu$1 values if gNaP is
totally removed from the soma and dendrite. 

In conclusion, the work described in this paper provides a novel way
to concisely quantify the fundamental nonlinearities underlying of
individual neurons. By allowing rigorous comparison of any neuronal
model with the behavior of real neurons, it makes possible to show
that nonlinear responses in voltage clamp are dominated by the active
dendritic structure. A determination of the molecular basis of the
eigenvalue analysis should provide a better understanding of how neurons
use their remarkable nonlinear properties in information processing.



\clearpage{}

\end{document}